\documentclass[a4paper]{jpconf}
\usepackage{graphicx}
\usepackage[labelfont=bf]{caption}
\usepackage{subcaption}
\usepackage{dcolumn}
\usepackage{bm}
\usepackage[labelsep=period]{caption}

\usepackage{xcolor} 

\usepackage{array}

\begin{document}
\title{Ultrastable lasers: 
	investigations of crystalline mirrors 
	and closed cycle cooling at 124 K}


\author{C Y Ma$^1$, J Yu$^1$, T Legero $^1$, S Herbers$^1$, D Nicolodi$^1$, M Kempkes$^1$, F Riehle$^1$, D Kedar$^2$, J M Robinson$^2$, J Ye$^2$ and U Sterr$^1$}

\address{$^1$ Physikalisch-Technische Bundesanstalt (PTB), Bundesallee 100, Braunschweig, Germany}
\address{$^2$ JILA, National Institute of Standards and Technology and University of Colorado, Boulder, CO, USA}

\ead{chun.ma@ptb.de}

\begin{abstract}

We have investigated crystalline AlGaAs/GaAs optical coatings with three ultra-stable cavities operating at 4 K, 16 K, 124 K and 297 K. 
The response of the cavities' resonance frequencies to variations in optical power indicates non-thermal effects beyond the photo-thermo-optic effect observed in dielectric coatings. 
These effects are strongly dependent on the intensity of the intracavity light at 1.5~\textmu m.
When the rear side of the mirrors is illuminated with external light, we observe a prominent photo-modified birefringence for photon energies above the GaAs bandgap, which points to a possible mechanism relating our observations to the semiconductor properties of the coatings.
Separately, we also present a low maintenance evolution of our 124 K silicon cavity system where the liquid nitrogen based cooling system is replaced with closed cycle cooling from a pulse-tube cryo-cooler.


\end{abstract}

\section{Introduction}
Lasers with long coherence time and narrow linewidth are an essential tool for quantum sensors and clocks.
Since the previous $8^{th}$ Symposium on Frequency Standards and Metrology in 2015, the field of ultrastable lasers has developed significantly. 
Ultrastable cavities and laser systems are now commercially available with fractional frequency instabilities in the mid $10^{-16}$ range.  
Ultrastable frequencies can now be distributed both locally and over long distances \cite{sch22a}.
With femtosecond combs, frequency noise of lasers at different wavelengths can be compared, allowing frequency stability transferred from one wavelength to another \cite{nic14,ben19, li22a} with added fluctuations well below $10^{-17}$ and to generate microwaves with exceptionally low phase noise \cite{xie17}. 
This enables interrogation times of optical clocks longer than a second \cite{ori18,cam17}.
Cavity-stabilized lasers have become very reliable and can even replace masers as flywheels to bridge downtimes of  optical clocks to create a fully optical and continuous timescale \cite{mil19}.

\begin{figure}[h]
	\begin{center}
		\includegraphics[width=0.6 \textwidth]{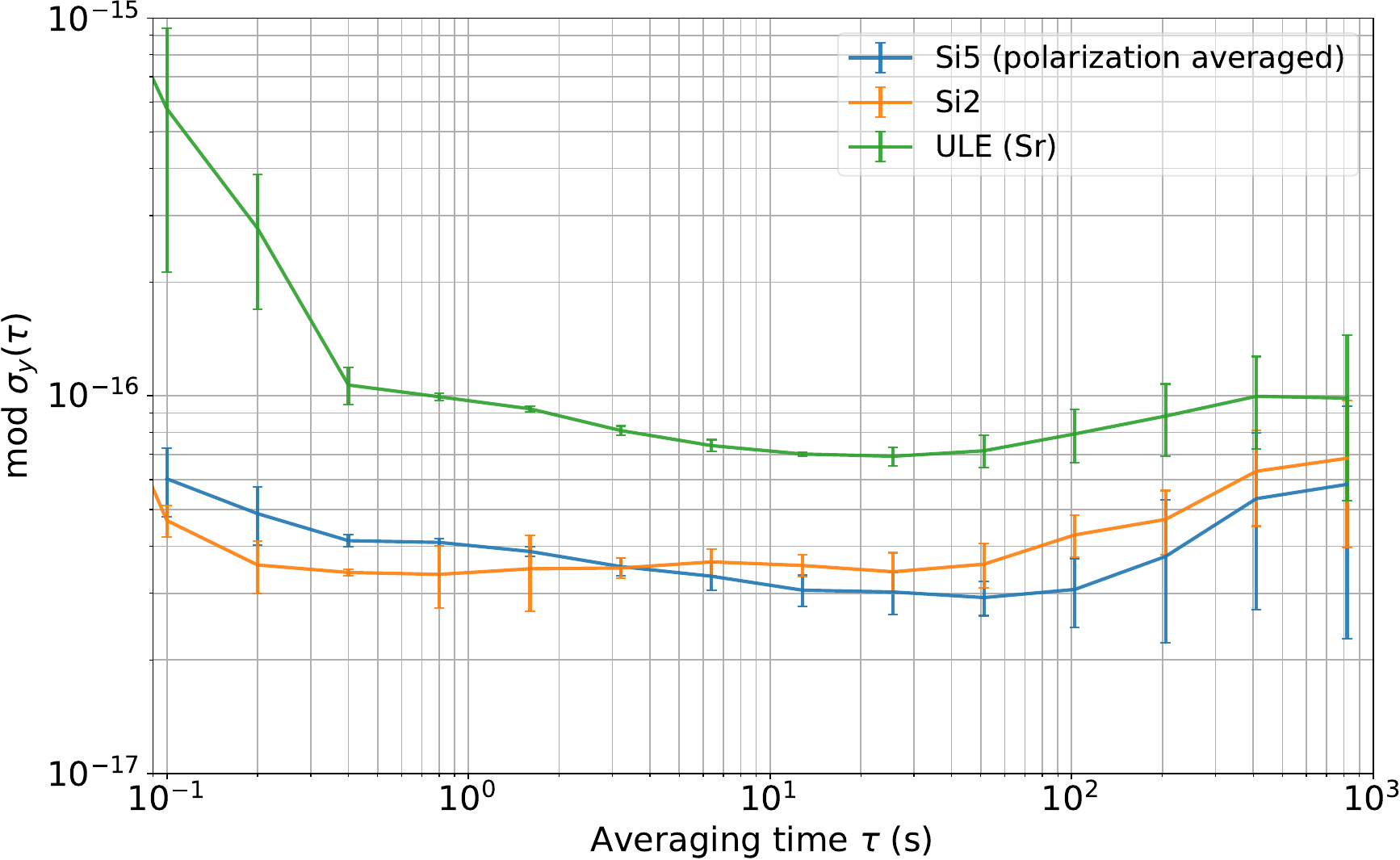}
	\end{center}
	\caption{
		Fractional frequency instability, expressed as modified Allan deviation, of a cryogenic silicon cavity with dielectric mirrors (Si2), 
		of an otherwise identical cavity with crystalline coatings (Si5), 
		and a 48 cm long room temperature cavity ULE (Sr) with dielectric coatings at 698 nm \cite{hae15a},
		determined from a three-cornered hat analysis. 
		The uncertainty bars indicate the standard deviation of ten separate evaluations.
	}
	\label{fig:TCH}
\end{figure}

For both terrestrial and space-borne transportable clocks, many transportable cavity designs and setups have been developed, the best reaching $1.6 \times 10^{-16}$ instability \cite{her22}.  
Laser instabilities as low as $7 \times 10^{-17}$ and $4 \times 10^{-17}$ have been reported for room temperature and cryogenic cavities respectively \cite{sch22a,mat17a}. 
Recently, we have measured the stability of our two cryogenic silicon cavities (Si2 and Si5) (see section 2) and of a 48 cm long ULE cavity at 698 nm \cite{hae15a} with the three-cornered-hat method \cite{yu23}. 
We obtained instabilities down to $3.2 \times 10^{-17}$ for the cryogenic silicon cavity with dielectric mirrors (Si2), whereas for an otherwise identical cavity with crystalline mirrors (Si5) an instability of $3.0 \times 10^{-17}$ at longer averaging times was found (see Fig. \ref{fig:TCH}).

Similar to large gravitational wave detectors \cite{har06b, gra20a}, the performance of these best sub-meter sized ultrastable optical resonators is currently limited by the Brownian fluctuations of the usually employed dielectric mirror coatings \cite{ked23,yu23a}. 
Brownian thermal noise is related to the mechanical loss by internal friction \cite{lev98} through the fluctuation-dissipation theorem \cite{cal51,kub66}.
Thus it can be reduced by employing optical coatings with lower mechanical loss.
Crystalline Al$_{0.92}$Ga$_{0.08}$As/GaAs Bragg reflectors \cite{col13,col23} have appeared as alternative to conventional {Ta$_2$O$_2$/SiO$_2$ dielectric multilayer coatings for reducing coating thermal noise. 
Their mechanical loss, which is about a factor of ten lower than conventional coatings, can reduce the Brownian noise, expressed in Allan deviation $\sigma_y$, by a factor of 3 \cite{col13}.
At the same time, a high reflectivity can be achieved, leading to a cavity finesse exceeding 300\,000 \cite{col16,ked24}.
Compared to dielectric coatings, these crystalline coatings exhibit much larger static birefringence on the order of 
$n_\mathrm{biref} \approx 7 \times 10^{-4}$ \cite{yu23a},
leading to frequency splittings between the linearly polarized fast and slow eigenmodes proportional to the free spectral range (FSR) of the cavity and $n_\mathrm{biref}$.

Our first studies at cryogenic temperatures showed novel noise contributions from the crystalline coating \cite{ked23,yu23a}.
After careful suppression of technical noise, remaining anticorrelated frequency fluctuations between the two eigenmodes that are dependent on intracavity power and mode area were observed. 
The anticorrelated fluctuations between the two polarization eigenmodes of this noise type relate to fluctuations of the coating birefringence, thus we refer to it as intrinsic birefringent noise. 
The Brownian thermal noise level was also determined and it agrees with the low mechanical loss value at room temperature. 
After removing the coating birefringent noise by simultaneously interrogating both modes, a remaining global noise in excess of the Brownian noise dominates the frequency instability at 124~K, 16~K and 4~K.

In addition to these intrinsic noises in AlGaAs coatings, previous studies at cryogenic temperatures have also observed an unexpected photo-induced frequency change \cite{yu23a}
beyond the photo-thermo-optic effect observed in dielectric coatings \cite{far12} arising from the temperature change caused by the laser power absorbed in the coatings. 
In the following sections we will focus 
on expanded investigations of the coating birefringence from cryogenic to room temperature. 
To gain more insight into potential physical mechanisms that might be related to the semiconductor properties of AlGaAs/GaAs, we also studied the response to changes in intracavity power and to broadband illumination at different wavelengths. 
Our findings may help to uncover the mechanisms related to the semiconductor properties and the noise of these coatings, thus potentially enabling a reduction of that noise.

\section{Experimental setup}

Three optical resonators operated at 1.5 $\mu$m were used to investigate the coating birefringence. 
Table \ref{tab:setups} summarizes their properties.
Two resonators employ monocrystalline silicon spacers with AlGaAs coatings bonded to silicon mirror substrates \cite{ked23,yu23a}.  
Based on our previous design \cite{hae15a}, the third resonator employs a 48 cm long ULE spacer and a pair of fused silica substrates coated with AlGaAs coatings. ULE rings are attached to the back of the mirrors to compensate for CTE mismatch \cite{leg10}, leading to a measured CTE zero crossing at 297 K. 

Similar to Si5 and Si6 \cite{ked23,yu23a}, the birefringence of the coatings in the ULE cavity results in a splitting of two polarization eigenmodes as shown in Table \ref{tab:setups}
with finesses of $1.290 (2) \times 10^{5}$ and  $1.282 (2) \times 10^{5}$ for slow and fast axes, respectively. 
To study the coating-specific optical length changes, two lasers are locked independently to this ULE cavity from opposite sides as in previous studies \cite{yu23a}. 
An optical heterodyne beat signal between the two lasers is detected from the transmitted and reflected beam at one side of the cavity, forming a common-path geometry to suppress the influence of path length fluctuations.

\begin{center}
	\begin{table}[h]
		\centering
		\caption{\label{tab:setups}
		Summary of three investigated ultrastable cavities with AlGaAs coatings. Mirror configuration is plane concave (pl/cc) or concave - concave (cc/cc) with the indicated radius of curvature (ROC) and the beam radius on the curved (Si cavities) or on the plane mirror (ULE cavity) is indicated. For Si5 and Si6 one mirror coating stack consists of 45.5 GaAs/AlGaAs layer pairs (total thickness 11.7 $\mu$m) whereas for ULE cavity the coating stack of 38.5 pairs  is 9.6 $\mu$m thick.
		}
		
		\begin{tabular}{c r r c c c c}
			\br
		    Cavity & Length & Temperature & Mirror          &   FSR   & Birefringent & Beam  \\
			         &        &             & configuration   &         & splitting    & radius\\
			\mr
			  Si5    & 21 cm  &  124 K      & cc/cc (2 m ROC) & 710 MHz & 200 kHz      & 482 $\mu$m\\
			  Si6    &  6 cm  & 4 K/16 K    & cc/cc (1 m ROC) & 2.5 GHz & 770 kHz      & 294 $\mu$m\\
		    ULE    & 48 cm  &  297 K      & pl/cc (1 m ROC) & 310 MHz & 104 kHz      & 495 $\mu$m\\ 
			\br
		\end{tabular}
	\end{table}
\end{center}

\section{Photo-induced effects}
It is well known that fluctuations of the intracavity power in optical cavities couple to frequency fluctuations via the photo-thermo-optic effect \cite{far12} and degrade the laser stability.
Stabilizing the intracavity power is thus critical. 

In addition to the photo-thermo-optic effect, for AlGaAs crystalline coatings, it has been observed that a change in intracavity optical  power also couples to changes in birefringence \cite{yu23a}. The mechanism for this coupling is not yet fully understood. To improve our understanding of this phenomenon,  we investigated the frequency change of three independent optical resonators locked to the fundamental Hermite-Gaussian HG$_{00}$ mode when the optical power is suddenly changed.

 
To account for the differences in finesse and mode size of the three cavities, we normalize our measurements to the average intensity $I$ at the AlGaAs coatings, as calculated from the intracavity power $P$ and the $1/e^2$ intensity beam radius $w$ 

\begin{equation}
	I = \frac{P}{\pi w^2}.
	\label{eq:I_av}
\end{equation} 


Fig. \ref{fig:step} shows the responses of the coating birefringence after a sudden increase of optical power. 
Si5 and Si6 are operated at temperatures with very low thermal expansion of silicon, thus the photo-thermal effect from the Si spacer and mirror substrates is negligible, 
and the induced frequency changes due to the modified birefringence are already visible in the measured frequency against a reference laser. 
For the ULE cavity at 297~K, the frequency difference between the fast and slow axes is calculated to remove the common photo-thermal contribution from fused silica substrates.
Half of the frequency difference is shown in the figure to make the results comparable to the other measurements.  

\begin{figure}[h]
	\begin{subfigure}[b]{0.50\textwidth}
		\includegraphics[width=18pc]{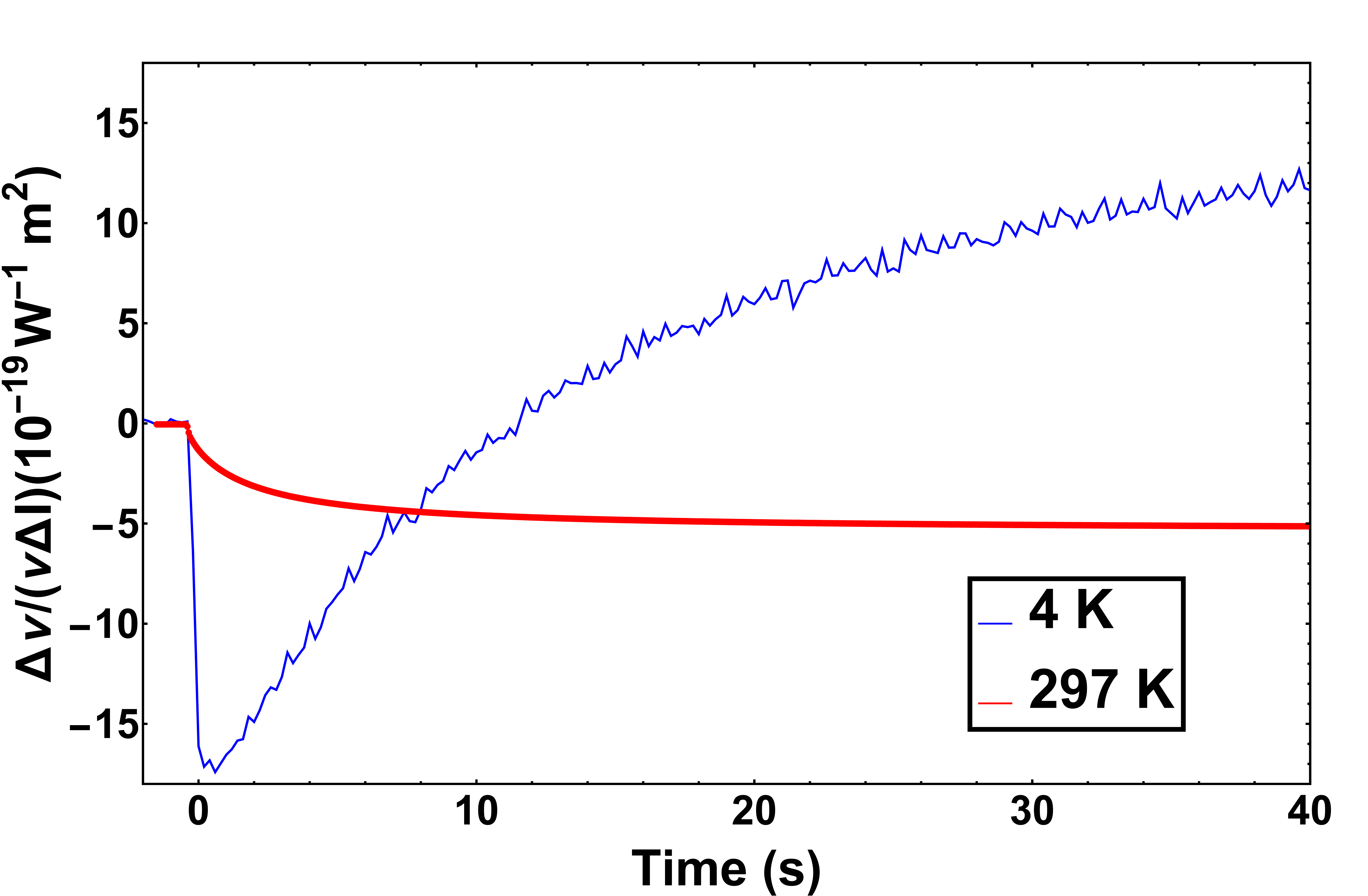}
	\end{subfigure}
	\begin{subfigure}[b]{0.50\textwidth}
		\includegraphics[width=18pc]{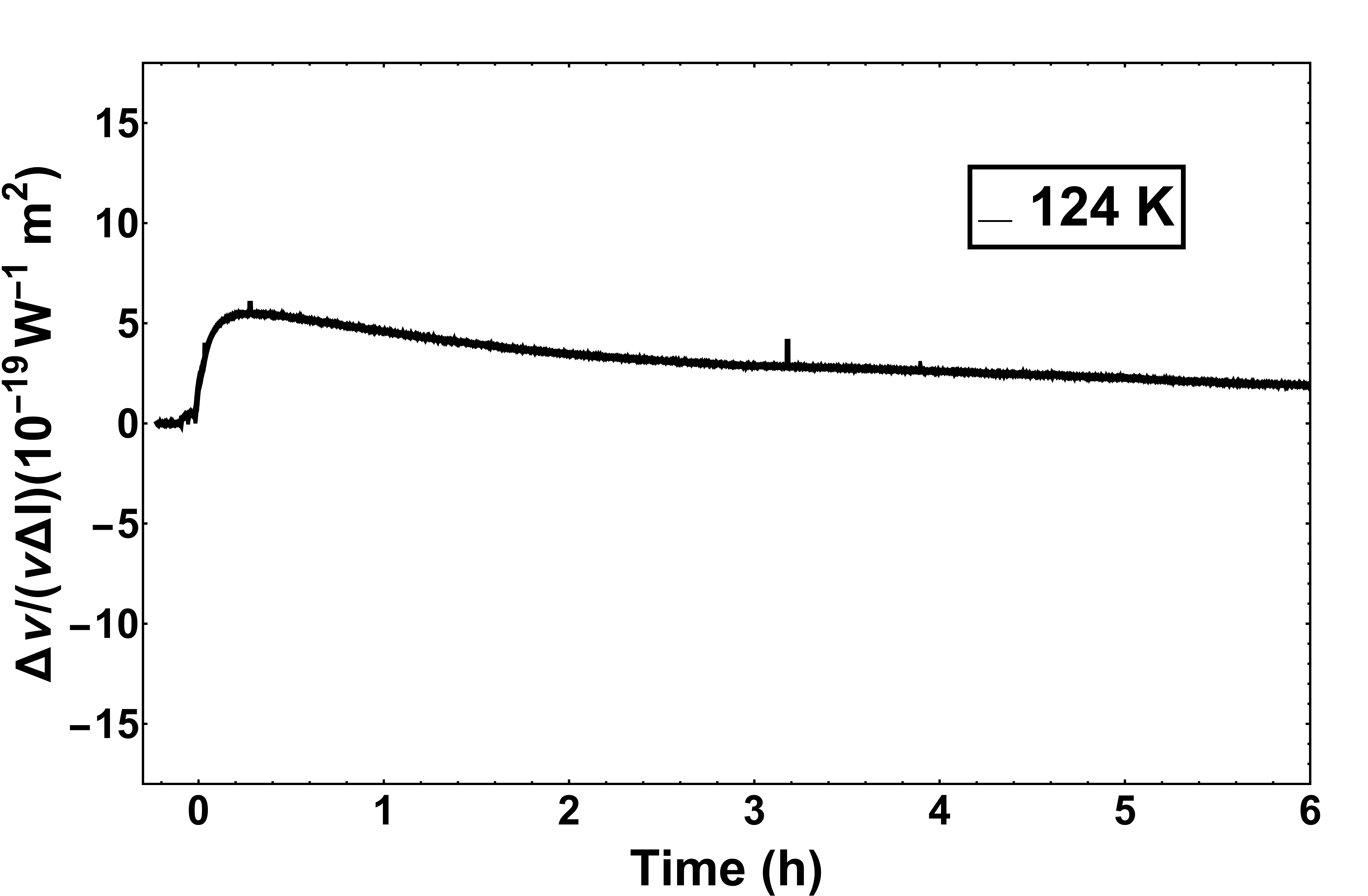}
	\end{subfigure}
	\caption{
		Fractional frequency change $\Delta \nu / \nu$ due to photo-modified birefringence after a sudden increase of intracavity power for the fast axis. 
		The fractional frequency change is normalized by the change in intensity $\Delta I$. 
		For the 4 K and 297 K measurement (left), the response is in the order of seconds, whereas for 124 K (right) the response is in the order of hours.
	}
	\label{fig:step}
\end{figure}

The observed change in frequency  $\Delta \nu_{\mathrm{photo}}$ 
can be related to the change of coating birefringence $\Delta n_{\mathrm{photo}}$ 

\begin{equation}
	\Delta \nu_\mathrm{photo} = \frac{\Delta n_\mathrm{photo}}{n_\mathrm{biref}}\nu_\mathrm{biref}
	\label{eq:d_n}
\end{equation} 

where ${n_\mathrm{biref}}$ is the static birefringence of the crystalline coatings in the individual cavities \cite{yu23a} and $\nu_{\mathrm{biref}}$ is the resulting line splitting. 
A schematic view of the intracavity light acting on the AlGaAs coating of the ULE cavity is shown in Fig. \ref{fig:Scheme}.

\begin{figure}[h]
	\begin{center}
		\includegraphics[width=0.45 \textwidth]{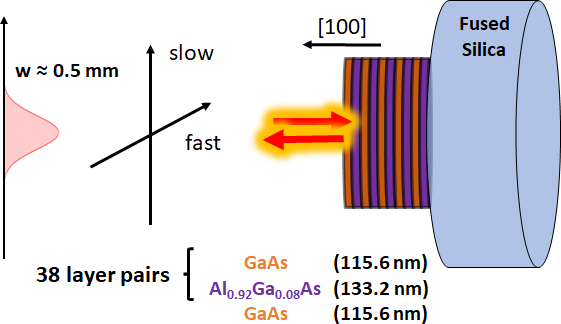}
	\end{center}
	\caption{
		Schematic view of the crystalline Bragg reflector bonded to fused silica substrates of the 297 K cavity. 
		The coating stack starts and ends with a GaAs layer.
	}
	\label{fig:Scheme}
\end{figure}

Taking the data as shown in Fig. \ref{fig:step}, the resulting change of birefringence $\Delta n_\mathrm{photo}/\Delta I$ of Si5 (124 K) is the smallest, which is followed by Si6 and the ULE cavity (297 K). 
The response at 124 K has a very different time scale compared to the responses at 4 and 297 K while the sign of the long term response at 297~K is different from the other two measurements. 

\subsection{Sensitivity to intracavity power} 

   
Fig. \ref{fig:intensity} (left) shows the line splitting as a function of the intracavity laser intensity on the mirror coatings measured with the ULE cavity. 
The range of intensity investigated corresponds to a transmitted power of 1.7~\textmu W to 108~\textmu W. 
Increasing intracavity intensity reduces the line splitting for the room temperature cavity. 
An opposite sign is observed in the 4~K and 124~K setups as well as in recent studies at room temperature \cite{zhu23, kra23a}.
This is possibly related to the fact that all setups besides our 297~K setup use ``flipped''  AlGaAs coatings \cite{col16}, where the [1 0 0] crystalline axis points in the opposite direction. 
The line splitting as function of intensity of our room temperature cavity is highly nonlinear, possibly indicating a saturation at high intensity. 
At the highest intracavity intensity level studied, 
the change of mirror birefringence $\Delta n_\mathrm{photo}$ is about 1\% of its static value ${n_\mathrm{biref}}$. 

\begin{figure}[h]
	\begin{subfigure}[b]{0.50\textwidth}
		\includegraphics[width=18pc]{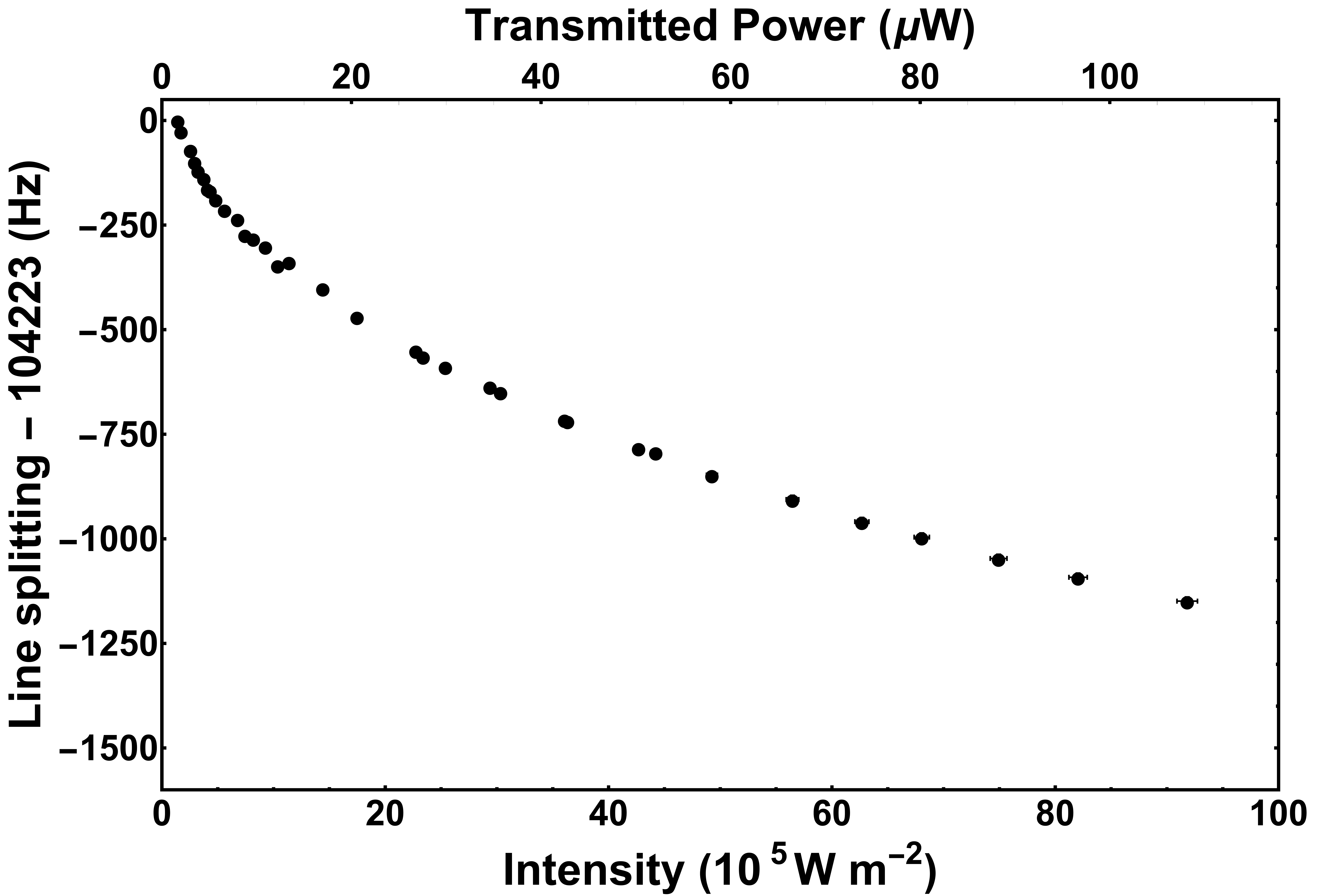}
	\end{subfigure}
	\begin{subfigure}[b]{0.50\textwidth}
		\includegraphics[width=18pc]{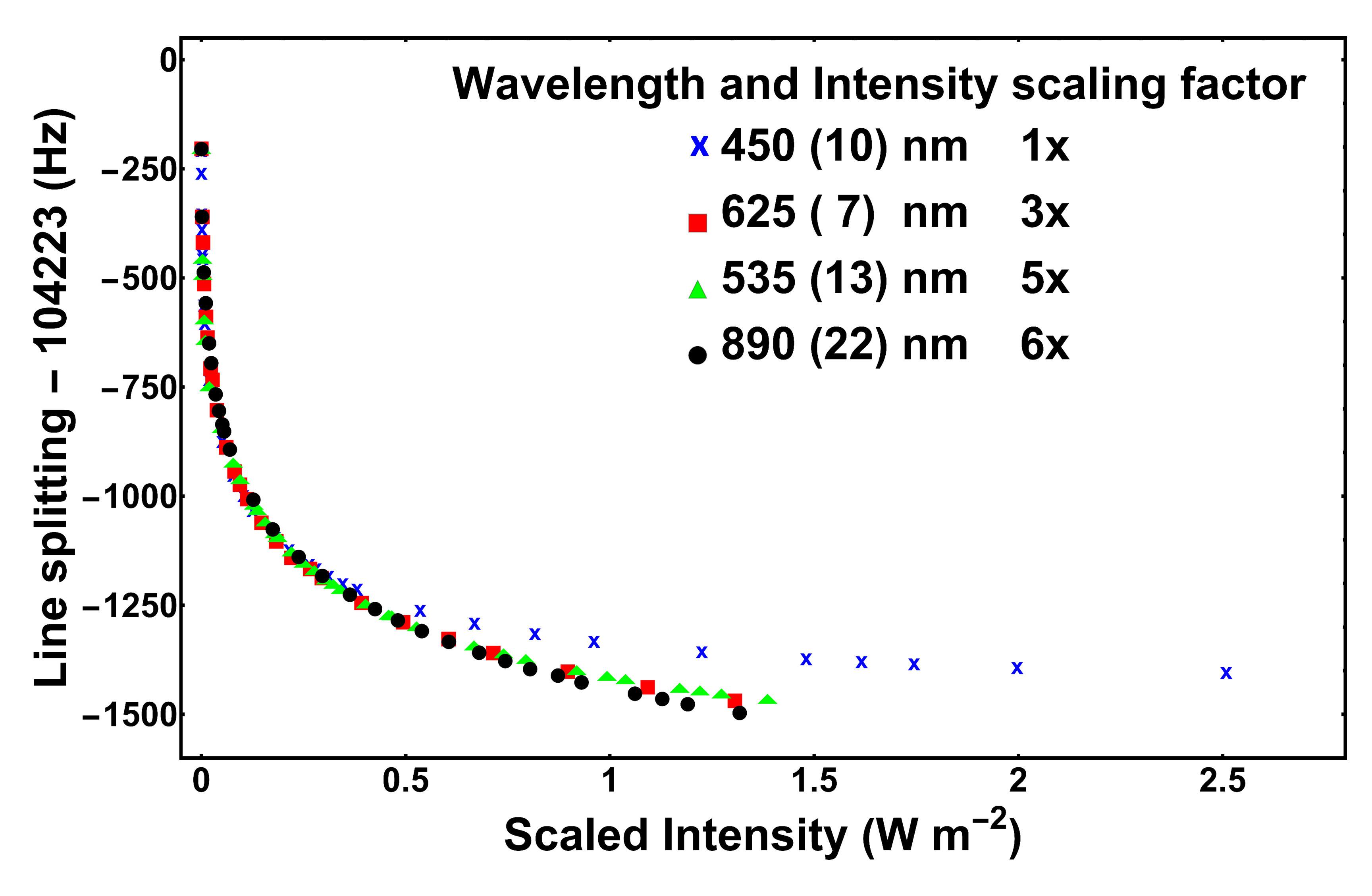}
	\end{subfigure}
	\caption{
	Birefringent splitting as a function of intracavity laser intensity (left) and external high power LED intensity (right) for the ULE cavity operating at room temperature after reaching steady state. 
	In the left graph the corresponding power transmitted by the cavity is also indicated on the top horizontal axis. 
	In the right graph the coating is illuminated by LEDs of four different wavelengths shorter than the GaAs bandgap at the same 1.5 $\mu$m intracavity laser intensity. 
	The curves can be superimposed by multiplying the intensity by a wavelength-dependent scaling factor. 
	A common frequency offset is subtracted for clarity, as indicated in the annotation.
	}
	\label{fig:intensity}
\end{figure}

\subsection{External illumination} 

So far the photo-induced effect has been investigated for intracavity light at 1.5~\textmu m. 
As the fused silica mirror substrates of the room temperature cavity are transparent to visible and near infrared light,
we illuminate the AlGaAs coatings from outside the vacuum chamber with LEDs to further investigate the wavelength dependence of $\Delta n_\mathrm{photo}$ above and below the bandgap of AlGaAs. 
The LEDs were calibrated to determine the intensity $I$ at the mirror coatings, accounting for divergence and the loss from the windows. 

The final line splittings with respect to the LED intensities are displayed in Fig. \ref{fig:intensity} (right). 
The response to the intensity is also highly nonlinear as seen in the response of intracavity power changes.
The responses at the different wavelengths can be described by a similar curve when their respective intensity is scaled by an empirically determined number.
The smallest sensitivity is observed for blue light (see Fig. \ref{fig:intensity} right). 
Comparing Fig. \ref{fig:intensity} left and right panels, the sensitivity of the line splitting over intensity is much larger for external illumination with LEDs and it also shows a larger curvature. 
Thus the line splitting response from the cavity light cannot be superimposed with the responses from the LEDs.

\begin{figure}[h]
	\begin{center}
		\includegraphics[width=0.65 \textwidth]{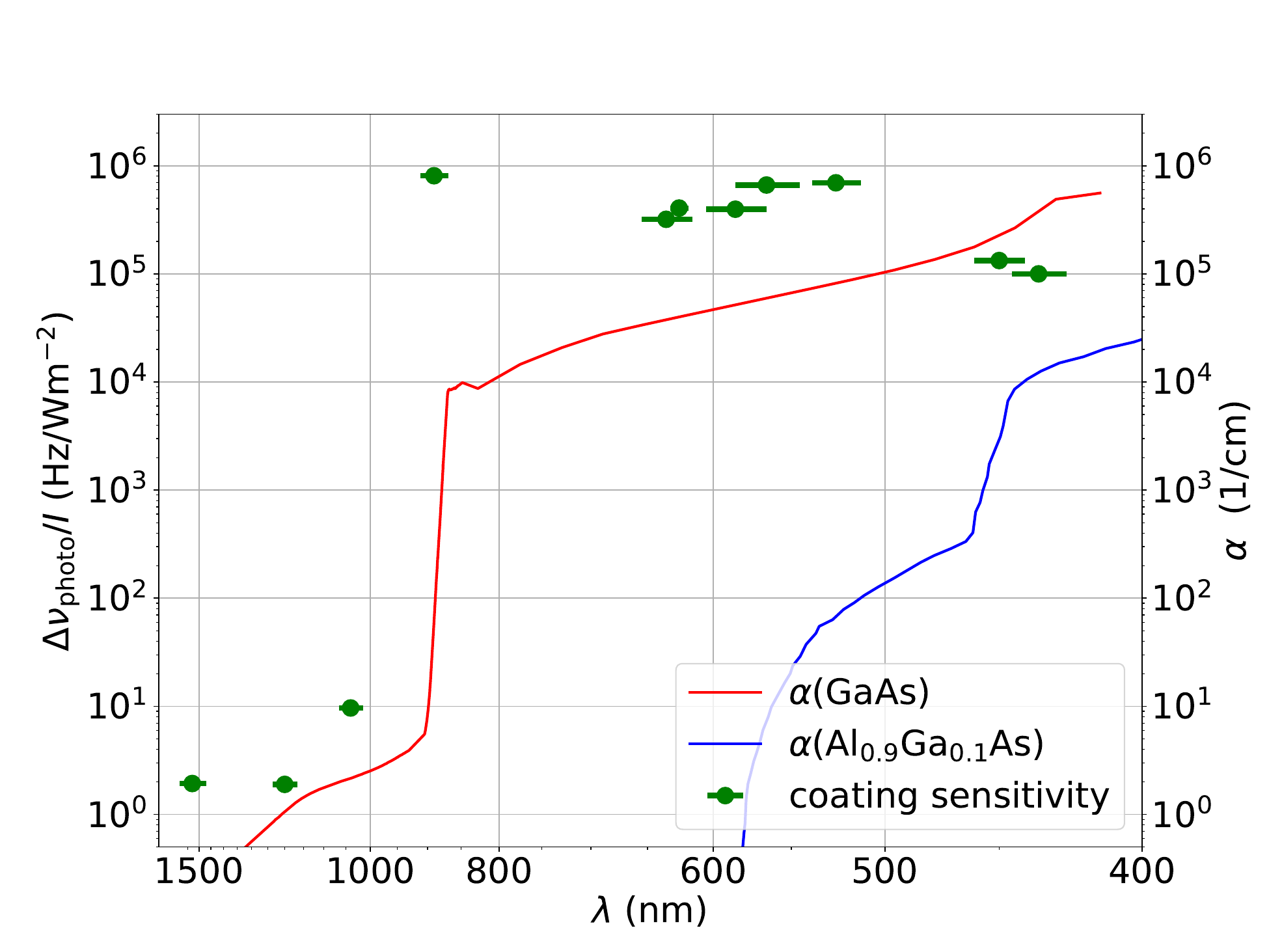}
	\end{center}
	\caption{Wavelength dependent sensitivity $\Delta \nu_\mathrm{photo}/I$ at low intensity in Hz/Wm$^{-2}$ (green dots) of photo-modified birefringence measured with narrowband LEDs. 
	The bandwidths of the LEDs are indicated by the horizontal bars.
	The uncertainty of the sensitivity is less than the marker size.
	The absorption coefficients $\alpha$ of GaAs (red) and  Al$_{0.9}$Ga$_{0.1}$As (blue) in cm$^{-1}$ are included for comparison \cite{mon76,stu62,hut08}.}
	\label{fig:sensitivity}
\end{figure}

Fig. \ref{fig:sensitivity} displays the observed sensitivity of the birefringent splitting to external light in Hz/Wm$^{-2}$ over a broader range of wavelengths between 450 nm and 1.5~\textmu m. To account for the nonlinearity in sensitivity, the largest sensitivity of each LED is shown.
The absorption profiles of GaAs and AlGaAs are included for comparison \cite{mon76,stu62,hut08}. 
The sensitivity drops by five orders of magnitude across the GaAs bandgap between 890 nm to 1050 nm, 
as for longer wavelengths no direct photoabsorption can occur in pure GaAs. 
For photon energies above the GaAs bandgap we observed a slight decrease in the sensitivity compared to the increase of the absorption coefficients towards shorter wavelengths, 
which might be due to the variation of penetration depth with wavelength. 
Blue light is already absorbed in the first GaAs layer in comparison to light at 890 nm that has an absorption length beyond ten layers of GaAs. 
This results in charges being created in different regions. 

\subsection{Simple model}
This wavelength-dependent behavior with correspondence to the absorption of GaAs indicates that photo-excitation of free charge carriers is likely to be the primary effect upon illumination. 
The carriers that are created in the first few~\textmu m thick absorption region on the back of the mirror coating stack migrate in the GaAs/AlGaAs heterostructure and finally create an electric field at the top layer. This modifies the birefringence through the electro-optic effect.
For an electric field perpendicular to the mirror surface, the linear electro-optic effect in GaAs creates a birefringence $\Delta n_\mathrm{EO}$ with the observed orientation and a sensitivity of \cite{nam61,tan23}
\begin{equation}
	\Delta n_\mathrm{EO} = r_{41} n_0^3 E
	\label{eq:EO}
\end{equation} 
with the electro-optic tensor component $r_{41} = -1.33$~pm/V and the GaAs refractive index $n_0 = 3.48$. 
Thus, the observed maximum $\Delta n_\mathrm{photo}$ of $8.7 \times 10^{-6}$  would require an electric field of 155 kV/m corresponding to 17 mV over the thickness of the top GaAs layer.

\subsection{Temporal behavior}

Previous investigation at cryogenic temperatures revealed a faster response at high final intensity \cite{yu23a}.
To understand the mechanism of the photo-induced effect in a better way, the temporal change of line splitting due to a step in intracavity power and external light are further studied at room temperature. 

Fig. \ref{fig:temporal} (left) shows three independent step responses of the splitting, when the optical power is stepped up or down to the same final level. 
As in Fig. \ref{fig:step}, the fractional frequency change is normalized by the change in optical intensity, considering the signs of the quantities. 
Due to the nonlinear sensitivity of the line splitting with intensity (as seen in Fig. \ref{fig:intensity}), the normalized frequency changes were further scaled (with scaling factors of 0.8, 1 and 1.09) to make the temporal change in frequency of all measurements overlap.
The good overlap indicates that the dynamical behavior is independent on the initial condition and the step size of intensity level. 

Similarly, the temporal response of the photo modified birefringence to external light is studied when LEDs at different wavelengths are switched on (Fig. \ref{fig:temporal} right).
The intensities of the LEDs were adjusted to result in the same birefringent frequency change $\Delta \nu_\mathrm{photo}$ = 300 Hz.
We observe the same temporal behavior, irrespective of the LED wavelength.
This observation is also in agreement with our simple model, based on initial charge creation by photoabsorption and charge migration which is followed by the modified birefringence via the electro-optic effect \cite{ked23,yu23a,tan23}.

\begin{figure}[h]
	\begin{subfigure}[b]{0.50\textwidth}
		\includegraphics[width=\textwidth]{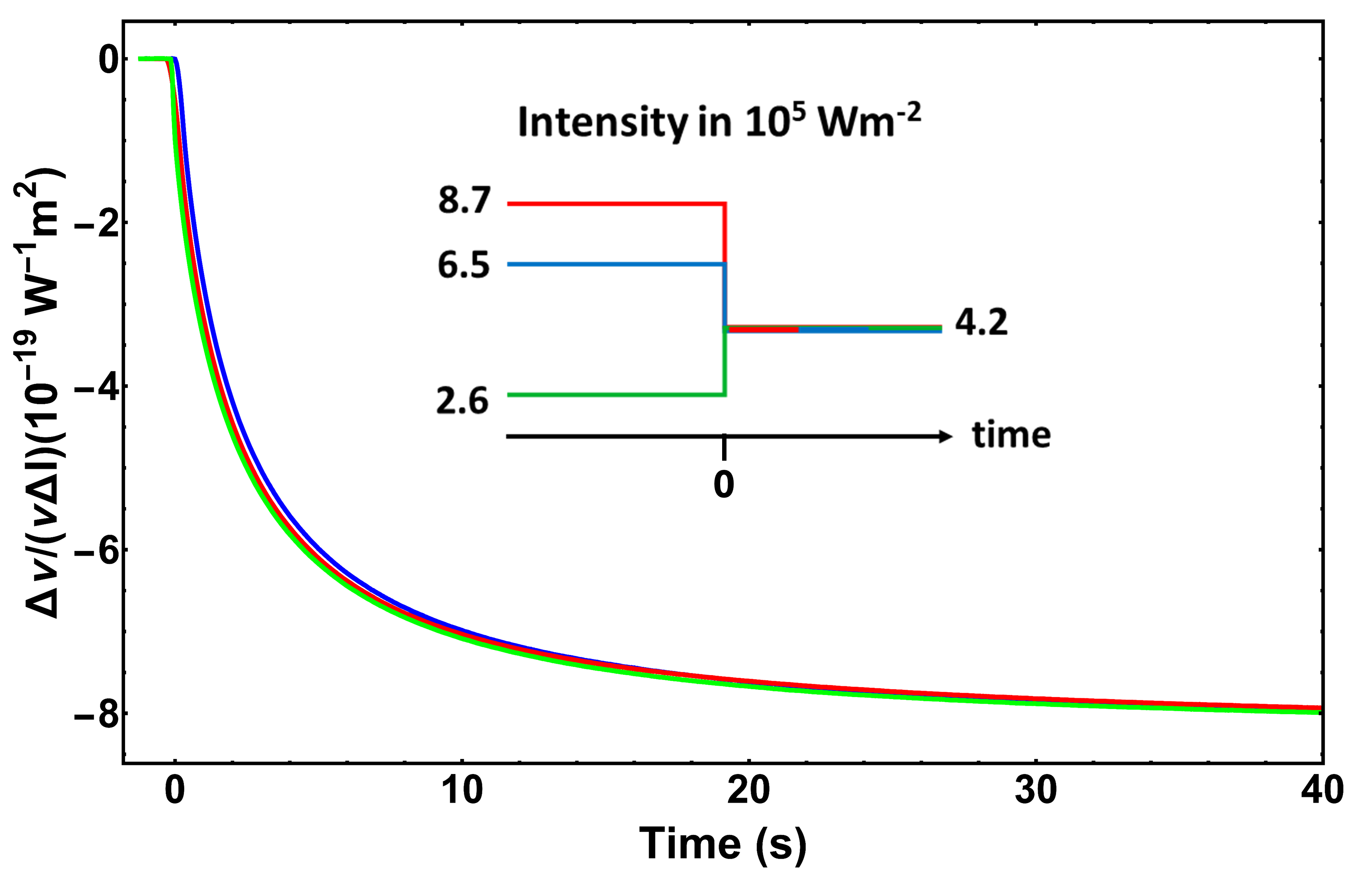}
	\end{subfigure}
	\begin{subfigure}[b]{0.50\textwidth}
		\includegraphics[width=\textwidth]{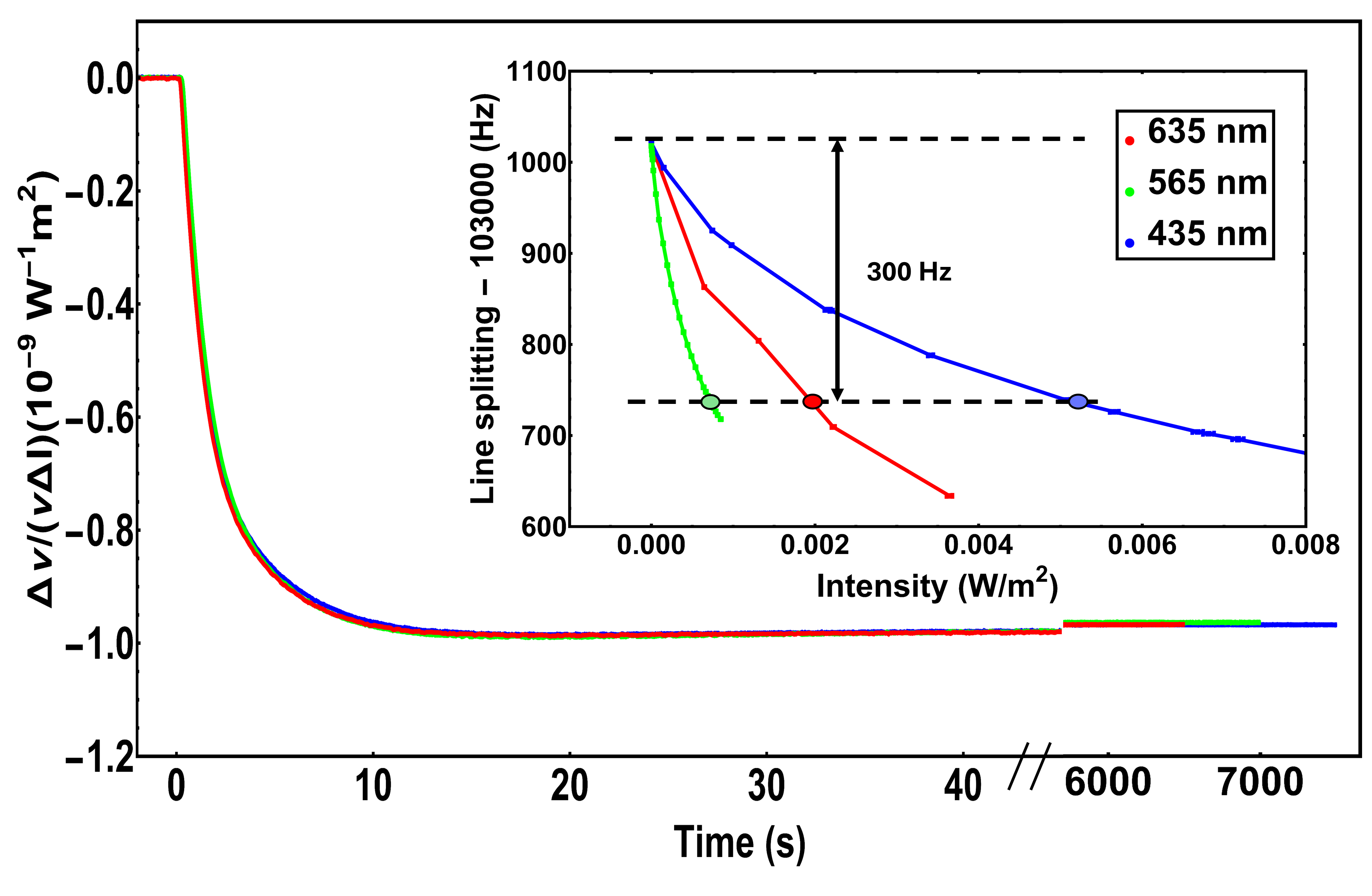}
	\end{subfigure}\hspace{2pc}%
	\caption{
	Temporal response of the birefringent splitting by intracavity light for different step sizes of intensities to the same final value (left). The colors of the curves correspond to the line colors of the steps in the inset.
	Response on switching on LEDs of different wavelengths (right). 
	The intensity (determined from the current) is chosen to get the same $\Delta \nu_\mathrm{photo}$ of 300 Hz (see inset). 
	}
	\label{fig:temporal}
\end{figure}

\section{Closed cycle cooling at 124~K}

Besides frequency stability, reliability and robustness are also important aspects for ultrastable lasers. 
In particular, the low-noise cooling of our current 124~K systems rely on cold nitrogen gas evaporated from a liquid nitrogen dewar \cite{kes12a, hag13a}. 
The continuous cooling of each Si cavity at 124~K requires 400 liters liquid nitrogen per week and frequent refillings. 
To simplify the operation and as a first step toward a transportable cryogenic system, we have replaced the liquid nitrogen based cooling system for one of our cryogenic cavity setups with a system where the cooling power is provided by a single-stage pulse-tube cryo-cooler. 
The cavity environment is cooled by a flow of cold helium gas that circulates between a heat exchanger connected to a pulse-tube cold-head and another heat exchanger connected to the outermost radiation shield in the vacuum chamber (Fig. \ref{fig:cryocooler}). 
As in the previous setup, vacuum-isolated flexible hoses are used to isolate the cavity from vibrations of the pulse-tube cooler.
The temperature of the radiation shield is actively controlled by modulating the rotation speed of the cryogenic fan used to maintain helium flow.
The cavity can be kept at 124~K operating the cryo-cooler with a cooling power of roughly 35~W at 85~K.
The system demonstrates sub-millikelvin outer shield temperature stability similar to the one obtained with the liquid nitrogen based cooling system.
A study is underway to characterize the impact of vibrations on the stability of the cavity resonance frequency. 

\begin{figure}[h]
	\begin{center}
		\includegraphics[width=0.75 \textwidth]{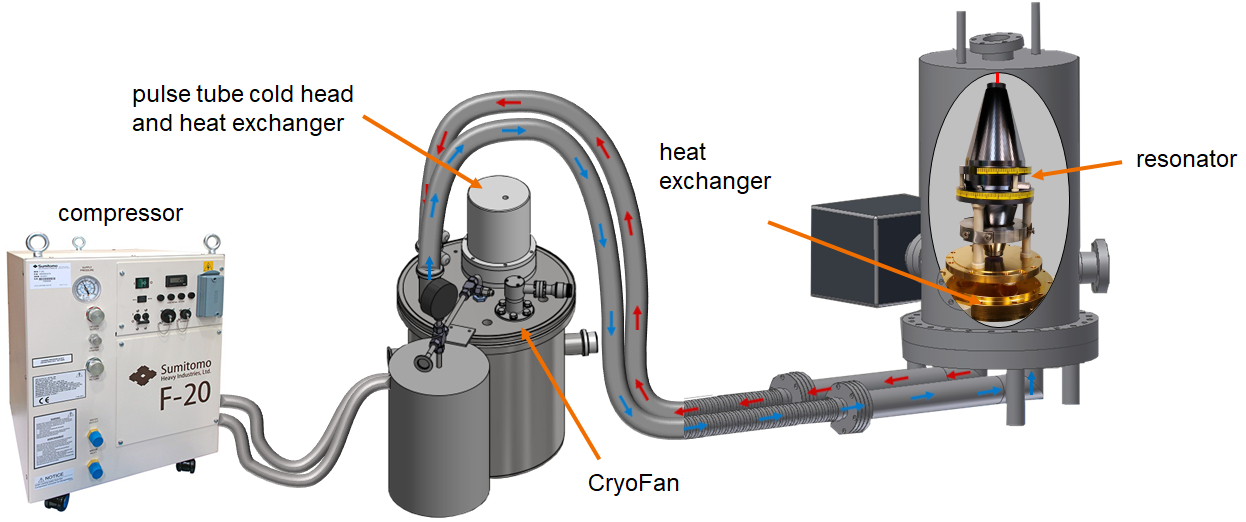}
	\end{center}
	\caption{Setup for a low-maintenance cryogenic 124 K Si cavity (not drawn to scale).}
	\label{fig:cryocooler}
\end{figure}

\section{Conclusions and outlook}

We have investigated the photo-induced effect on the birefringence of crystalline coatings at 4~K, 16~K, 124~K and 297 K. 
Upon a sudden change in optical power, the original birefringence is modified at all temperatures and changes to a new stationary value. 
At room temperature, the change of birefringence with optical power was observed to reduce at high intracavity intensity. 
This non-thermal response at 297~K at different wavelength was also studied for the first time by illuminating the coating with external LED light in a broad range of wavelengths. 
At photon energies above the GaAs bandgap, the birefringence is modified five orders of magnitude more than at energies below the bandgap. 
The temporal behavior of intracavity and external light reveals the dynamic change in birefringence is dependent on the extent of the modified birefringence due to a different photo-excitation rate. 
We explain the observed effect with initial photoabsorption creating charge carriers.  These charges migrate and create an electric field.
The resulting field induces an electro-optic effect on the coatings that modifies the birefringence. 

From the measured sensitivities, a relative power stabilization of 10$^{-6}$ is required for a photo-induced instability below $1 \times 10^{-17}$ at room temperature. 
As next step, the coating associated noise contributions with and without external light will be investigated to determine the utmost performance of AlGaAs coatings at room temperature. 

\ack
We acknowledge support by the Project 20FUN08 NEXTLASERS, which has received funding from the EMPIR programme cofinanced by the Participating States and from the European Union’s Horizon 2020 Research and Innovation Programme, and by the Deutsche Forschungsgemeinschaft (DFG, German Research Foundation) under Germany’s Excellence Strategy–EX-2123 QuantumFrontiers (Project No. 390837967), SFB 1227 DQ-mat (Project No. 274200144). 
This work is partially supported by the Max Planck-RIKEN-PTB Center for Time, Constants and Fundamental Symmetries. 
This work is also supported by NIST, National Science Foundation QLCI OMA-2016244, AFRL, and JILA Physics Frontier Center (NSF Grant No. PHY-1734006).

\section*{References}
\bibliographystyle{iopart-num}
\bibliography{texbiFSM}

\providecommand{\newblock}{}
\begin{thebibliography}{10}
\expandafter\ifx\csname url\endcsname\relax
  \def\url#1{{\tt #1}}\fi
\expandafter\ifx\csname urlprefix\endcsname\relax\def\urlprefix{URL }\fi
\providecommand{\eprint}[2][]{\url{#2}}

\bibitem{sch22a}
Schioppo M, Kronj\"ager J, Silva A, Ilieva R, Paterson J~W, Baynham C~F~A,
  Bowden W, Hill I~R, Hobson R, Vianello A, Dovale-\'Alvarez M, Williams R~A,
  Marra G, Margolis H~S, Amy-Klein A, Lopez O, Cantin E, \'Alvarez-Mart\'inez
  H, Le~Targat R, Pottie P~E, Quintin N, Legero T, H\"afner S, Sterr U, Schwarz
  R, D\"orscher S, Lisdat C, Koke S, Kuhl A, Waterholter T, Benkler E and
  Grosche G 2022 {\em Nature Commun.\/} {\bf 13} 212

\bibitem{nic14}
Nicolodi D, Argence B, Zhang W, Le~Targat R, Santarelli G and Le~Coq Y 2014
  {\em Nature Photonics\/} {\bf 8} 219--223

\bibitem{ben19}
Benkler E, Lipphardt B, Puppe T, Wilk R, Rohde F and Sterr U 2019 {\em Opt.
  Express\/} {\bf 27} 36886--36902 also see erratum \cite{ben20}

\bibitem{li22a}
Li P, Rolland A, Jiang J and Fermann M~E 2022 {\em Opt. Express\/} {\bf 30}
  22957--22962

\bibitem{xie17}
Xie X, Bouchand R, Nicolodi D, Giunta M, H\"ansel W, Lezius M, Joshi A, Datta
  S, Alexandre C, Lours M, Tremblin P~A, Santarelli G, Holzwarth R and Le~Coq Y
  2017 {\em Nature Photonics\/} {\bf 11} 44--47

\bibitem{ori18}
Origlia S, Pramod M~S, Schiller S, Singh Y, Bongs K, Schwarz R, Al-Masoudi A,
  D\"orscher S, Herbers S, H\"afner S, Sterr U and Lisdat C 2018 {\em Phys.
  Rev. A\/} {\bf 98} 053443

\bibitem{cam17}
Campbell S~L, Hutson R~B, Marti G~E, Goban A, Darkwah~Oppong N, McNally R~L,
  Sonderhouse L, Robinson J~M, Zhang W, Bloom B~J and Ye J 2017 {\em Science\/}
  {\bf 358} 90--94

\bibitem{mil19}
Milner W~R, Robinson J~M, Kennedy C~J, Bothwell T, Kedar D, Matei D~G, Legero
  T, Sterr U, Riehle F, Leopardi H, Fortier T~M, Sherman J~A, Levine J, Yao J,
  Ye J and Oelker E 2019 {\em Phys. Rev. Lett.\/} {\bf 123}(17) 173201

\bibitem{her22}
Herbers S, H\"{a}fner S, D\"{o}rscher S, L\"{u}cke T, Sterr U and Lisdat C 2022
  {\em Opt. Lett.\/} {\bf 47} 5441--5444

\bibitem{mat17a}
Matei D~G, Legero T, H\"afner S, Grebing C, Weyrich R, Zhang W, Sonderhouse L,
  Robinson J~M, Ye J, Riehle F and Sterr U 2017 {\em Phys. Rev. Lett.\/} {\bf
  118} 263202

\bibitem{hae15a}
H{\"a}fner S, Falke S, Grebing C, Vogt S, Legero T, Merimaa M, Lisdat C and
  Sterr U 2015 {\em Opt. Lett.\/} {\bf 40} 2112--2115

\bibitem{yu23}
Yu J 2023 {\em Cryogenic silicon {Fabry}-{Perot} resonator with
  {Al$_{0.92}$Ga$_{0.08}$As/GaAs} mirror coatings.\/} Ph.D. thesis
  QUEST-Leibniz-Forschungsschule der Gottfried Wilhelm Leibniz Universit\"at
  Hannover

\bibitem{har06b}
Harry G~M, Armandula H, Black E, Crooks D~R~M, Cagnoli G, Hough J, Murray P,
  Reid S, Rowan S, Sneddon P, Fejer M~M, Route R and Penn S~D 2006 {\em Appl.
  Opt.\/} {\bf 45} 1569--1574

\bibitem{gra20a}
Granata M, Amato A, Cagnoli G, Coulon M, Degallaix J, Forest D, Mereni L,
  Michel C, Pinard L, Sassolas B and Teillon J 2020 {\em Appl. Opt.\/} {\bf 59}
  A229--A235

\bibitem{ked23}
Kedar D, Yu J, Oelker E, Staron A, Milner W~R, Robinson J~M, Legero T, Riehle
  F, Sterr U and Ye J 2023 {\em Optica\/} {\bf 10} 464--470

\bibitem{yu23a}
Yu J, H\"afner S, Legero T, Herbers S, Nicolodi D, Ma C~Y, Riehle F, Sterr U,
  Kedar D, Robinson J~M, Oelker E and Ye J 2023 {\em Phys. Rev. X\/} {\bf
  13}(4) 041002

\bibitem{lev98}
Levin Y 1998 {\em Phys. Rev. D\/} {\bf 57} 659--663

\bibitem{cal51}
Callen H~B and Welton T~A 1951 {\em Phys. Rev.\/} {\bf 83} 34--40

\bibitem{kub66}
Kubo R 1966 {\em Rep. Prog. Phys.\/} {\bf 29} 255--284

\bibitem{col13}
Cole G~D, Zhang W, Martin M~J, Ye J and Aspelmeyer M 2013 {\em Nature
  Photonics\/} {\bf 7} 644--650

\bibitem{col23}
Cole G~D, Ballmer S, Billingsley G, Cata\~no Lopez S~B, Fejer M, Fritschel P,
  Gretarsson A~M, Harry G~M, Kedar D, Legero T, Makarem C, Penn S~D, Reitze D,
  Steinlechner J, Sterr U, Tanioka S, Truong G~W, Ye J and Yu J 2023 {\em Appl.
  Phys. Lett.\/} {\bf 122} 110502

\bibitem{col16}
Cole G~D, Zhang W, Bjork B~J, Follman D, Heu P, Deutsch C, Sonderhouse L,
  Robinson J, Franz C, Alexandrovski A, Notcutt M, Heckl O~H, Ye J and
  Aspelmeyer M 2016 {\em Optica\/} {\bf 3} 647--656

\bibitem{ked24}
Kedar D, Yao Z, Ryger I, Hall J~L and Ye J 2024 {\em Optica\/} {\bf 11} 58--63

\bibitem{far12}
Farsi A, Siciliani~de Cumis M, Marino F and Marin F 2012 {\em J. Appl. Phys.\/}
  {\bf 111} 043101

\bibitem{leg10}
Legero T, Kessler T and Sterr U 2010 {\em J. Opt. Soc. Am. B\/} {\bf 27}
  914--919

\bibitem{zhu23}
Zhu X~Q, Cui X~Y, Kong D~Q, Yu H~W, Jiang X, Xu P, Dai H~N, Chen Y~A and Pan
  J~W 2023 {\em Fourteenth International Conference on Information Optics and
  Photonics (CIOP 2023)\/} vol 12935 ed Yang Y International Society for Optics
  and Photonics (SPIE) p 1293541

\bibitem{kra23a}
Kraus B 2023 {\em A highly stable {UV} clock laser\/} Ph.D. thesis
  QUEST-Leibniz-Forschungsschule der Gottfried Wilhelm Leibniz Universit\"at
  Hannover

\bibitem{mon76}
Monemar B, Shih K~K and Pettit G~D 1976 {\em J. Appl. Phys.\/} {\bf 47}
  2604--2613

\bibitem{stu62}
Sturge M~D 1962 {\em Phys. Rev.\/} {\bf 127}(3) 768--773

\bibitem{hut08}
Hutchby J~A and Fudurich R~L 2008 {\em J. Appl. Phys.\/} {\bf 47} 3140--3151

\bibitem{nam61}
Namba S 1961 {\em J. Opt. Soc. Am.\/} {\bf 51} 76--79

\bibitem{tan23}
Tanioka S, Vander-Hyde D, Cole G~D, Penn S~D and Ballmer S~W 2023 {\em Phys.
  Rev. D\/} {\bf 107}(2) 022003

\bibitem{kes12a}
Kessler T, Hagemann C, Grebing C, Legero T, Sterr U, Riehle F, Martin M~J, Chen
  L and Ye J 2012 {\em Nature Photonics\/} {\bf 6} 687--692

\bibitem{hag13a}
Hagemann C 2013 {\em Ultra-stable laser based on a cryogenic single-crystal
  silicon cavity\/} Ph.D. thesis Fakult\"at f\"ur Mathematik und Physik der
  Gottfried Wilhelm Leibniz Universit\"at Hannover
  \urlprefix\url{http://edok01.tib.uni-hannover.de/edoks/e01dh13/739734164.pdf}

\bibitem{ben20}
Benkler E, Lipphardt B, Puppe T, Wilk R, Rohde F and Sterr U 2020 {\em Opt.
  Express\/} {\bf 28} 15023--15024

\end{thebibliography}
\end{document}